# FirstPersonScience: Quantifying Psychophysics for First Person Shooter Tasks

Josef Spjut, NVIDIA

Ben Boudaoud, NVIDIA

Kamran Binaee, NVIDIA and Rochester Institute of Technology

Zander Majercik, NVIDIA

Morgan McGuire, NVIDIA

Joohwan Kim, NVIDIA

ABSTRACT

*In the emerging field of esports research, there is an increasing demand for quantitative results that can be used by players, coaches and analysts to make decisions and present meaningful commentary for spectators. We present FirstPersonScience, a software application intended to fill this need in the esports community by allowing scientists to design carefully controlled experiments and capture accurate results in the First Person Shooter esports genre. An experiment designer can control a variety of parameters including target motion, weapon configuration, 3D scene, frame rate, and latency. Furthermore, we validate this application through careful end-to-end latency analysis and provide a case study showing how it can be used to demonstrate the training effect of one user given repeated task performance.*

## Introduction

In competitive First Person Shooter (FPS) video games, players are frequently faced with a need to hit their opponent before their opponent hits them. The player to fire first while maintaining enough accuracy to hit the target will win any engagement that is otherwise strategically equivalent. Carefully controlled user studies are needed to understand what game and hardware characteristics are most important for player performance in competitive FPS games. FirstPersonScience is a tool we developed to conduct these kinds of user studies. Using FirstPersonScience, researchers can design user studies and gather high quality objective measurements of player behavior and system performance while isolating variables of interest (Kim, et al., 2019). We provide FirstPersonScience as an open source project to enable other researchers to conduct similar studies and make it more feasible to reproduce scientific results.

## Features

FirstPersonScience supports a variety of features for psychophysical analysis of users, game software design exploration and computer hardware analysis. These three pillars are central to the sort of user studies we believe will be most useful.

### *Psychophysics*

Psychophysics is the scientific study of the relation between stimulus and sensation (Gescheider, 2013). The common experimental paradigm in psychophysics is to present a series of stimuli to human participants and analyze the responses to understand sensation. The stimulus presentation scheme, tasks given to subjects, and analysis methods have been tailored to minimize or account for behavioral noise that can occur during data collection. Thus, we recommend adopting the methods of psychophysics to accurately measure performance level in FPS tasks. At a high level, the software design of FirstPersonScience enables rapid iteration of experiment design as seen in Figure 1.

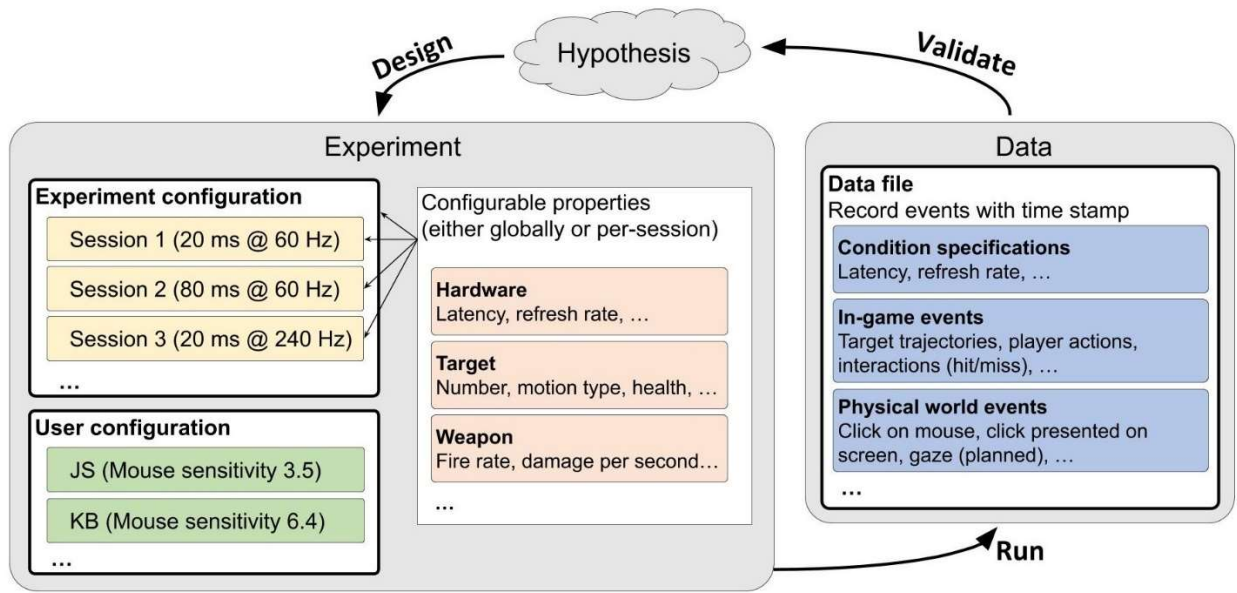

Figure 1: FirstPersonScience allows experiment designers to configure an experiment, gather results, and iterate on the experimental design rapidly.

To facilitate adoption, we include two stimulus level management schemes of psychophysics in FirstPersonScience: method of constant stimuli and staircase. In the method of constant stimuli, the experimenter pre-defines stimulus levels for each condition. These stimulus levels are randomly selected and provided to the subject. In the staircase method, the program adjusts stimulus level adaptively based on subject's responses. This enables a more efficient estimation of a subject's performance level. At the moment, the experimenter can use the method of constant stimuli through text-based configuration, and an experimenter can specify either of the two schemes for each condition with source code modifications, after which our library manages the stimulus level and data collection procedure for the programmed conditions. In the future, we may add text-based configuration support for the staircase stimulus method.

*Game Software*

FirstPersonScience provides a minimal FPS-like interface, employing a simple, centered crosshair/reticle based aiming system wherein the player view direction is controlled using the mouse. It is similar in concept to many FPS trainer applications available today (Aimlab, 2019) (KovaaK's FPS Aim Trainer, 2019). We used G3D (McGuire, Mara, & Majercik, 2017) as the core game engine for development, and it is required to build FirstPersonScience from source. We have not yet produced a standalone executable, but one may be produced in the future which could be used without access to source code modification without a G3D installation. The 3D scene is selectable from any supported format, and standard G3D scene formats are usable (see the following section). The player can be prevented from moving within the scene once spawned if the experimental design calls for it, or movement can be allowed for other kinds of experiments. A number of standard game rendering features are supported, including screen space ambient occlusion (Ritschel, Grosch, & Seidel, 2009), temporal antialiasing (Karis, 2014), and others. An example screen shot of the in-game view can be seen in Figure 2.

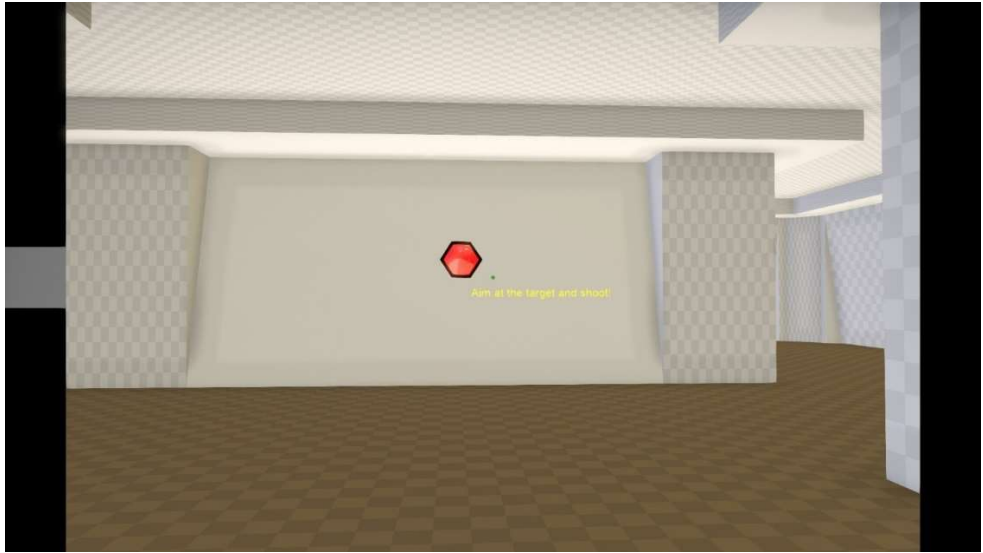

Figure 2: Example in-game view including tutorial instructions to player.

We designed the mouse motion to mimic that of commercial FPS games. Support for a third person camera view could be added by including an offset for the camera's center of rotation, in which case a character model will also need to be added. We adopted distance per 360° turn[1] as our mouse sensitivity metric to easily match a player's settings from other applications when combined with a sensitivity calculator such as one of the many available on the web (Mouse Sensitivity Calculator, 2019). We calibrated our distance per 360° scale by taking multiple measurements manually and iterating until the distance matched within a couple millimeters for a target of 30 cm per 360° rotation. This was necessary because G3D uses its own mouse sensitivity parameter that does not directly match commercial game engines.

User configurations are specified in a file that allows multiple users to be listed, each of which can set their own distance per 360° and mouse Dots Per Inch (DPI) as well as an identifier for that user that is recorded with the data collection. Another file, specific to a given experiment, keeps track of the status of each user for that experiment, recording which sessions have been completed. The experiment designer can also specify a per-user ordering of sessions to control for training effects in the experiment design.

*Experiment Configuration*

In order to set up the experiment design we provide a variety of configurable parameters that control in-game conditions. At a high level, sets of conditions are separated into *experiments*, which are in turn subdivided into *sessions* each containing sets of *trials* including the trial type and number to be administered for each type. Each trial represents a grouping of *target motion types* along with other specified experimental condition variables.

The experiment can be configured by setting a description, selecting a 3D scene to use, indicating which session ordering to use and specifying the weapon. The experiment then includes a list of target motions and a list of sessions which refer to the target motion types through sets of trials. The durations set at the experiment level include the time spent preparing for a trial (readyDuration), the time allocated for each task (taskDuration) and the time allowed for user feedback (feedbackDuration). Session ordering is done randomly, though serial ordering and other options will be added as needed for future experiments.

---

[1] Distance per 360-degree turn is the physical distance (in inches or centimeters) that the mouse must be moved to cause the player's view to rotate 360 degrees in game, returning to the starting view direction.

The weapon is configured by providing the amount of ammo to allow per trial, the fire period, damage per second of the weapon, and a flag to specify whether the weapon should continue to fire while the mouse button is held down or not. When the fire period is set below the frame period and auto fire is enabled, the weapon becomes like many "laser-style" weapons with continuous damage over time, similar to the primary fire weapon from the Zarya character in Overwatch. When the fire period is set to 0.5s with 6 ammo, the weapon is configured to more closely match the primary fire weapon from the McCree character in Overwatch. We provide a small set of suggested weapon configurations along with FirstPersonScience to help people get started. Additionally, the experiment designer can select weapon models to use, turn on or off miss decals and projectile drawing, and specify the sound to use when the weapon fires. In the future, support may be added for custom aiming reticles per weapon, controlling damage falloff with distance, adjusting bullet spread, and specifying decals to use for bullets that hit a wall.

The experiment configuration includes a list of target motion type specifications, which are each identified by a description. A target motion is specified by providing a minimum and maximum value for each parameter range, which are randomized within the valid range when the target is spawned. For static parameters, the minimum and maximum values can be matched. Target speed is randomized within its valid range when a time in the "motionChangePeriod" range has passed since the last speed and direction change. There is an option to lock the target to horizontal only motion, and an option to enable a ballistic jumping motion. An additional set of jump parameters control the speed, the period, and gravity for ballistic jumps. In the future, more kinds of target motion may be added.

To tie all these parameters together, a list of sessions is the final required configuration. Each session represents a single atomic that a user must complete in one sitting. Since sessions can be configured as "training" or "real", a typical methodology is to require one training session followed by the real session with similar settings. Since FirstPersonScience has been used to study frame rate and system latency, the experiment designer can specify a target frame rate for each session as well as the number of frames of delay to add in order to simulate configurations with higher latency. If the computer is powerful enough, it is possible to set the frame rate above the display refresh rate, in which case FirstPersonScience will complete more frames than the number that are shown on the display. This approach can be useful for further reducing the computer latency as is commonly done by esports competitors. The session then lists a set of trials, which specify the id of each target motion type to be used in that session as a reference to the target motion type specification discussed previously. The session executes the trials in random order, though other ordering options may be added as needed until the specified number of trials have been completed.

During development, the experiment designer can turn the "playMode" setting off to gain access to a developer mode, which includes several sliders and knobs to control various aspects of the application. Once the experiment is configured as desired, and the experiment is ready to run, "playMode" should be turned on, which removes the developer controls and sets the application to full screen mode to enable higher performance and remove distraction.

We use the ".Any" file type to specify our configurations, which is a superset of JSON and other types of test-based data serialization formats that are human readable. Sample configurations are provided with the application and are copied to user editable files on the first run. Since these files are human readable text, someone unfamiliar with programming can edit a few text fields and specify a new experiment design. Some user settings can also be adjusted within the application, such as mouse distance per 360°.

Validation
We validated our application using a mix of controlled tests and measurement. Since end-to-end latency and frame rate were among the first paramters targeted for experimentation, we heavily prioritized these parameters in validation. Through use of a mix of hardware measurement tools and software control/instrumentation we were able to carefully monitor and control sources of latency and assure our application was running as designed on the experiment machines.

*Hardware*
In order to help validate performance of the FirstPersonScience application we developed a simple, in-house tool for measuring click-to-photon latency in common mouse-to-monitor applications. Our tool uses an Arduino platform to sense the mouse button down event (from a hardware modified gaming mouse) and the resulting flash from the weapon firing on-screen using an amplified photodetector circuit. The resulting µs-accurate timestamped events can then be differenced to produce click-to-photon timing.

The platform was used through development of FirstPersonScience (as well as during studies) to validate and continuously monitor that actual click-to-photon latencies were in the expected range of values. This proved invaluable early in the design process when assumptions regarding correct configuration for low latency rendering approaches were still being developed. As an example, for a target of 60 fps, Figure 3 shows the distribution of click to display latencies as measured by our hardware tool. When possible, we recommend measuring photon latency at the vertical center of the display to get the average delay since there can be up to a full frame time in difference between the top of the display and the bottom, and the middle should give the average. The middle is also a more interesting location in FPS games because it is the usual position for the aiming crosshair. However, there may be situations where different parts of the screen are more important, in which case those areas of the screen should be measured.

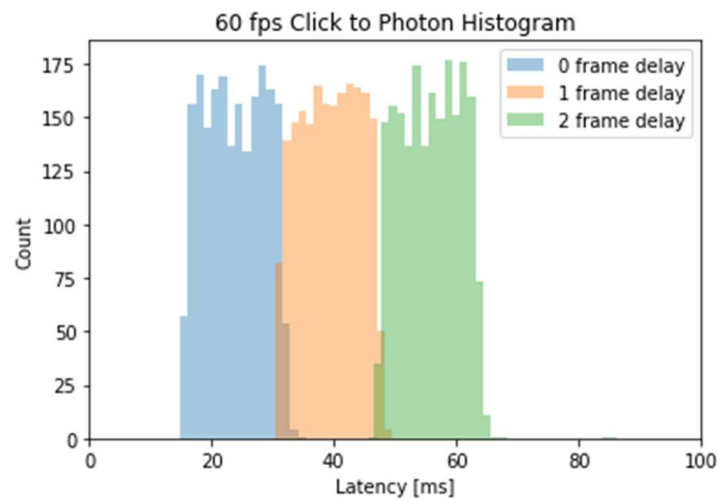

Figure 3: Measured click to photon latencies for 2000 clicks at 0, 1, and 2 frames of delay. Computer configured with Acer Predator XB252Q Monitor, NVIDIA RTX 2080 Ti, Intel Core i7-4790K, and Logitech G203 Mouse.

Demo

FirstPersonScience has been used previously to study the effects of refresh rate and latency independently (Kim, et al., 2019; Spjut, et al., 2019). In order to show the flexibility of this tool, we did a test to demonstrate the training effect as a single volunteer (one of the authors) continues to perform the same task over time. In this experiment, we configured FirstPersonScience in the 60 frames per second (fps) setting with 2 frames of delay, with the expected latency distribution as seen in Figure 3. A single target motion type was selected, in which the target would choose new speed and direction ever 1-2 seconds at random. The volunteer completed 60 trials, and a score was reported for every group of 10 trials. Across those 6 groups, the reported scores were 34, 43, 39, 41, 35, 48. While there was not a strict increase in score, a more general view of the data reveals a significant difference in task completion time for the first half of trials and the second half. In this experiment, some of the targets were challenging to hit, moving continually in one direction and causing the volunteer to miss enough times that the trial was considered at failure after 6 seconds (the configured trial duration). We removed those trials from the data for this analysis, leaving 55 successful trials which can be seen plotted in Figure 4.

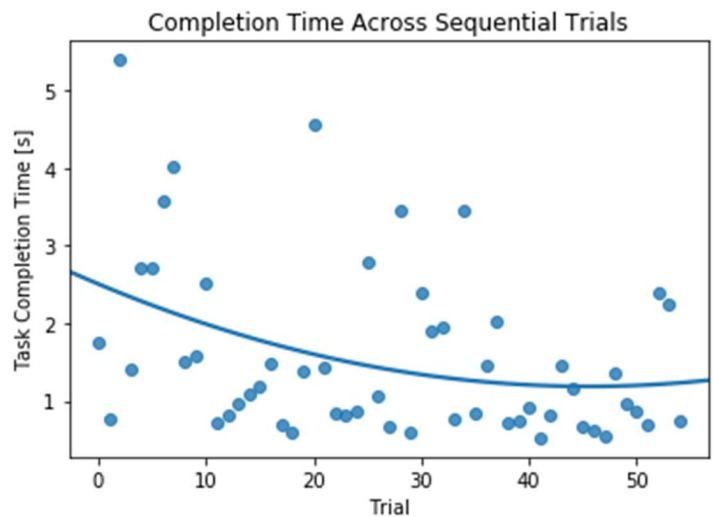

Figure 4: Scatter plot of task completion time across 55 trials with quadratic fit function demonstrating the training effect for a single user over a few minutes.

There is a general trend in this data of decreased task completion time, indicating the user improving at the task over the short time frame of a few trials. The first 27 trials resulted in a mean task completion time of 1.78 seconds with a standard error metric of 0.242 while the final 28 trials have a mean of 1.34 seconds with a standard error metric of 0.164. It is well known that a training effect is common to many tasks, including FPS tasks, thus we expect these results. A useful experimental methodology, therefore, may be to include an initial set of trials dedicated to allowing the user to train for the task and condition in question, after which the real experimental data can be gathered. If necessary, the training session can be used to map out the training trajectory, after which the expected training effect can be removed from final results.

Future Work

In the future, we plan to expand the functionality of FirstPersonScience as needed for the range of experiments that would be interesting to the esports community. Additional methods of variable exploration may be added to the psychophysical tool set, and more weapon features could be included. As discussed previously, a 3rd person camera offset would be of interest for 3rd person shooters such as Fortnite. Player and weapon models would enrich the gameplay, and weapon parameters would allow more diverse kinds of weapons to be represented and studied. One simple modification would be to add weapon clips with reload time as is common to many kinds of weapons in FPS games.

Conclusion

The field of esports research is growing rapidly, and a need is emerging for high quality scientific experimentation and research. FirstPersonScience is one tool among the spectrum of needed tools to enable scientific progress in understanding and modeling the high-performance human computer interfaces abundant in competitive esports. While FPS games are only one genre of the rich field of esports, we believe its popularity has made it a central focus, particularly as games like Fortnite and Overwatch develop serious league play and bring in high level competitors, coaches and team owners. While specific games will need carefully designed experiments to study individual quirks and interactions, FirstPersonScience gives us the ability to study FPS gameplay and tasks in the abstract, discovering the foundations of esports science. We hope that this tool will be useful for the esports research community to conduct meaningful user studies and gather data that shed more light on which aspects of the system contribute meaningfully to player performance.